\documentclass{acm_proc_article-sp}
\usepackage{algorithm}
\usepackage{algorithmic}
\usepackage{xfrac}
\usepackage{enumerate}
\usepackage{graphicx}
\usepackage{amsmath}

\begin{document}
\widowpenalty=100000
\clubpenalty=100000
\raggedbottom

% --- Author Metadata here ---
%\conferenceinfo{KDD}{'14 New York, New York, USA}
%\CopyrightYear{2007} % Allows default copyright year (20XX) to be over-ridden - IF NEED BE.
%\crdata{0-12345-67-8/90/01}  % Allows default copyright data (0-89791-88-6/97/05) to be over-ridden - IF NEED BE.

\title{The path most travelled: Mining road usage patterns from massive call data}
\numberofauthors{5} 
\author{
% 1st. author
\alignauthor
Jameson L. Toole\\
       \affaddr{Engineering Systems Division }\\
       \affaddr{Massachusetts Institute of Technology}\\
       \affaddr{Cambridge MA, USA}\\
       \email{jltoole@mit.eu}
% 2nd. author
\alignauthor
Serdar \c{C}olak\titlenote{Corresponding author}\\
       \affaddr{Civil and Environmental Engineering}\\
      \affaddr{Massachusetts Institute of Technology}\\
       \affaddr{Cambridge MA, USA}\\
       \email{serdarc@mit.edu}
% 3rd. author
\alignauthor 
Fahad Alhasoun\\
       \affaddr{Center of Computational Engineering}\\
      \affaddr{Massachusetts Institute of Technology}\\
       \affaddr{Cambridge MA, USA}\\
       \email{fha@mit.edu}
\and  % use '\and' if you need 'another row' of author names
% 4th. author
\alignauthor 
Alexandre Evsukoff\\
       \affaddr{School of Applied Mathematics}\\
       \affaddr{Get\'{u}lio Vargas Foundation}\\
       \affaddr{Rio de Janeiro, Brazil}\\
       \email{alexandre.evsukoff@fgv.br}
%% 5th. author
%\alignauthor 
%Nelson F. F. Ebecken\\
%       \affaddr{Civil Engineering Department}\\
%       \affaddr{Federal University of Rio de Janeiro}\\
%       \affaddr{Rio de Janeiro, Brazil}\\
%       \email{nelson@ntt.ufrj.br}
% 6th. author
\alignauthor
Marta C. Gonz\'{a}lez \\
       \affaddr{Civil and Environmental Engineering}\\
      \affaddr{Massachusetts Institute of Technology}\\
       \affaddr{Cambridge MA, USA}\\
       \email{martag@mit.edu}
}

\date{\today}

\maketitle
\begin{abstract}
Rapid urbanization places increasing stress on already burdened transportation systems, resulting in delays and poor levels of service. Billions of spatiotemporal call detail records  (CDRs) collected from mobile devices create new opportunities to quantify and solve these problems. However, there is a need for tools to map new data onto existing transportation infrastructure. In this work, we propose a system that leverages this data to identify patterns in road usage. First, we develop an algorithm to mine billions of calls and learn location transition probabilities of callers. These transition probabilities are then upscaled with demographic data to estimate origin-destination (OD) flows of residents between any two intersections of a city. Next, we implement a distributed incremental traffic assignment algorithm to route these flows on road networks and estimate congestion and level of service for each roadway. From this assignment, we construct a bipartite usage network by connecting census tracts to the roads used by their inhabitants. Comparing the topologies of the physical road network and bipartite usage network allows us to classify each road's role in a city's transportation network and detect causes of local bottlenecks. Finally, we demonstrate an interactive, web-based visualization platform that allows researchers, policymakers, and drivers to explore road congestion and usage in a new dimension. To demonstrate the flexibility of this system, we perform these analyses in multiple cities across the globe with diverse geographical and sociodemographic qualities. This platform provides a foundation to build congestion mitigation solutions and generate new insights into urban mobility.
\end{abstract}

\category{J.4}{Social and Behavioral Sciences}[Economics, Sociology]


%\keywords{ACM proceedings, \LaTeX, text tagging}

\section{Introduction} \label{Intro}
According the United Nations Population Fund (UNFPA), 2008 marked the first year in which the majority of the planet's population lived in cities.  Urbanization, already over 80\% in many western regions, is increasing rapidly as migration into cities continue. This rapid growth places enormous strain on infrastructures at a time when resources and funding remain scarce. Transportation systems, critical in providing residents with access to places, people, and goods, quickly become seized with congestion. Delays and poor levels of service not only waste time, money, and energy, but are also detrimental to social welfare as shown by recent research discovering the strong correlation between socioeconomic and geographic mobility \cite{chetty2014land,gurley2005effects, glaeser2008poor}.

In parallel to rapid urbanization, ubiquitous mobile computing, namely the pervasive use of cellular phones, has generated a wealth of data that can be analyzed to understand and improve urban infrastructure systems.  Calls, tweets, and other digital crumbs of millions of users are passively recorded along with meta-data containing geographic and social information. While there are very real and important privacy concerns raised by the use of this data, there are also countless new opportunities to mine patterns of human behavior and develop new strategies to make our cities more efficient, sustainable, and fun. Raw data alone, however, is not enough to solve the problems currently faced by urban transportation systems. We must discover the proper algorithms and metrics that combine multiple data sources to generate insights and knowledge about the use of city infrastructure so that we can provide policy makers or consumers actionable information. In doing so, we base our paper on three distinct areas of research specifically subjecting mobility and cities. 

Making use of this new massive data, a body of work has shed light on human mobility patterns, finding individuals predictable, unique, and slow to explore new places \cite{gonzalez2008understanding, brockmann2006scaling, de2013unique, song2010limits, song2010modelling}. Surprisingly, many of these properties appear to translate across different cities and continents despite differences in culture, socioeconomic variables, and geography. The benefits of this data have been realized in various contexts such as mobility motifs \cite{schneider2013unravelling, sevtsuk2010does}, disease spreading \cite{belik2011natural, wesolowski2012quantifying} and population movement\cite{lu2012predictability}. While these works have laid an important foundation, there still is a need to address the problem of mapping these inferred mobility patterns onto existing transportation infrastructure\cite{wang2012understanding}.

Another body of work has looked extensively at the physical and statistical properties of cities and the road infrastructure itself. Optimal transport in network models has been thoroughly studied \cite{carmi2008transport,li2010towards,wu2006transport} as well as the real cities and road systems. The geographic and network properties of road systems and land parcels are the result of emergence in self-organized complex systems and often recoverable \cite{batty2008size, barthelemy2011spatial, barthelemy2008modeling, lammer2006scaling}. These studies, however, rarely include any discussion of who uses this infrastructure and how that uses varies by time of day or day of week or by location.

Finally, the transportation and urban planning communities have a deep tradition in estimating and analyzing the movement of people through cities via various transportation systems. Arguably the most central puzzle piece in methods developed by transportation planners is the origin-destination information, abbreviated as \textit{OD} throughout this work.  OD information looks to estimate the number of trips between any two points in the city at a given time. It has traditionally been obtained through a combination of meticulous methods of statistical sampling \cite{daganzo1980optimal, smith1979design} and national household travel surveys \cite{stopher2007household,richardson1995survey}. While the tools and techniques developed by this community provide an excellent framework to think about mobility, they are simply not designed to integrate massive new data sources. The incompatibility between current methods and sheer volume and variety of data now available creates a gap in research where we base our paper. %Current work has begun to update these tools for new data and computing requirements, finding new patterns and regularities related to road usage along with important leverage points for policy makers to maximize system improvements with scarce resources.  

Our aim in this paper is to continue in those goals by presenting a system to mine billions of mobile phone traces and transforming them into estimates of road usage patterns. We begin by framing the built system in section \ref{architecture}. We explain our methods of extracting, cleaning, and storing road network information in section \ref{roadnets}. In section \ref{odgeneration}, we continue by detailing an algorithm to learn location transition probabilities from mobile phone call sequences. We then spatially join these calls with census data using a computationally efficient discretization procedure and upscale these probabilities into flows of vehicles. Section \ref{assignment} explains our implementation of a fast distributed incremental traffic assignment algorithm to map vehicle flows onto road networks.  We then move on to analyze transportation system performance and usage utilizing our inferred demand in section \ref{vvoc} and explore the relationship between census regions and road segments in section \ref{bipartite}. In section \ref{roadclass}, we classify road segments based on metrics that relate to their topological importance and the extent of usage. We demonstrate benchmarking results in section \ref{benchmark}. In section \ref{visualization}, we elaborate our interactive visualization platform aimed at better conveying our findings and helping to provide this information to policy makers and consumers through a compelling and easily understandable medium. Finally we end with our conclusions in section 4.  To demonstrate the flexibility of our system, we perform these analyses for five metro regions spanning countries and cultures: Boston, USA, San Francisco, USA, Rio de Janeiro, Brazil, Lisbon, Portugal, and Porto, Portugal. 

\subsection{List of Contributions\label{contributions}}
In this work we extend and scale previous work in five ways:
\begin{enumerate}
\item We describe a procedure to generate OD matrices by combining CDR data with population and demographic data provided by survey and census.
\item We parse OpenStreetMaps (OSM) -- inferring various street properties such as number of lanes, speed, and capacity -- to create routable and comparable road networks in many cities currently lacking such data.
\item We implement a fast, distributed incremental traffic assignment algorithm to route millions of trips made between intersections in cities in seconds.
\item We perform a comparison of congestion and road usage patterns in multiple cities with different geographies and cultures.
\item We provide an interactive visualization platform for use by researchers, citizens, and policy makers to better inform their decisions regarding transportation and mobility.
\end{enumerate}

\subsection{Description of Data\label{data}}
A key challenge in extracting mobility patterns and consequently estimating road usage lies in the integration of a variety of data sources. Location data gathered from devices such as mobile phones, while massive, is often noisy and tends to be biased due to differences in the usage of the technology and penetration rates amongst different segments of the population. Moreover, call detail records provide irregular location sampling frequency, as they only include location information when a call or text is made, and provide no insight into the mode of transportation used. To correct for these factors, data obtained from the census is used to scale the measured number of mobile phone trips by the phone market share  and the average vehicle usage rate of residents in that area (e.g. those living in downtown Boston use vehicles for roughly just 20\% of their trips versus almost 100\% for suburban residents). Each call is be spatially joined with the corresponding census tract associated with the residence of the mobile phone user, which creates an additional computational challenge.  Finally, detailed road networks are obtained and cleaned to ensure that they are topologically correct and incorporate realistic speed, length, and capacity profiles so that expected travel times can be estimated for the obtained routes.

To perform these tasks, our system combines and processes the following data sources:

\begin{description}
\item {\it Call Detail Records (CDRs):} At least three weeks of call detail records from mobile phone use across the subject city. The data includes the timestamp and the location for every phone call (and in some cases SMS) made by all users of a particular carrier. The spatial granularity of the data varies between cell tower level where calls are mapped to towers and triangulated geographical coordinate pairs where each call has a unique pair of coordinates accurate to within a few hundred meters. Market shares associated with the carriers that provide the data also vary. Personal information is anonymized through the use of hashed identification strings. 
\item {\it Census Data:} At the census tract (or equivalent) scale, we obtain the population and vehicle usage rate of residents in that area.  For US cities, the American Community Survey provides this data on the level of census tracts (each containing roughly 5000 people).  Census data is obtained for Brazil through IBGE (Instituto Brasileiro de Geografia e Estat\'{i}stica) and for Portugal through the Instituto de Nacional de Estatistica. All cities analyzed in this work have varying spatial resolutions of the census information.
\item {\it Road Networks:}  For many cities in the US, detailed road networks are made available by local or state transportation authorities. These GIS shapefiles generally contain road characteristics such as speed limits, road capacities, number of lanes, and classifications.  Often, however, these properties are incomplete or missing entirely.  Moreover, as such road inventories are expensive to compile and maintain, they simply do not exist for many cities in the world. In this case, we turn to OpenStreetMaps (OSM), an open source community dedicated to mapping the world through community contributions.  For cities where a detailed road network cannot be obtained, we parse OSM files  and infer required road characteristics to build realistic and routable networks. 
\end{description}

Table \ref{T1} compiles descriptive statistics for these data sources for each city.

\begin{table}
\centering
\begin{tabular}{lccccc}
 & \multicolumn{5}{c}{City}  \\
\cline{2-6}
 & Bos & Bay & Rio & Lis & Por \\
\hline
Population (mil.) & 4.5 & 7.15 & 12.6 & 2.8 & 1.7 \\
Area ($1000 km^{2}$) & 4.6 & 18.1 & 43.6 & 2.9 & 2.0 \\
% Length of Data (week) &  &  &  \\
\# of Users (mil.) & 1.65 & 0.43 & 2.19 & 0.56 & 0.47\\
\# of Calls (mil.) & 905 & 429 & 1,045 & 50 & 33 \\
\# of cell towers & N/A & 892 & 1421 & 743 & 335 \\ 
\hline
\# of Edges (ths.) & 21.8 & 24.3 & 40.9 & 28.1 & 15.1 \\
\# of Nodes (ths.) & 9.6 & 11.3 & 22.1 & 16.1 & 8.6 \\
%\# of Roads & 21,859 & 24,349 & 40,990 & 28,152 & 15,141 \\
%\# of Intersections & 9,643 & 11,308 & 22,088 & 16,060 & 8,576 \\
\hline
\# of Tracts & 732 & 1139 & 381 & 295 & 272 \\
\hline
\end{tabular}
\caption{A comparison of the extent of the data involved in the analysis of the subject cities. ($Bos$: \textit{Boston, US}, $Bay$: \textit{San Francisco Bay Area, US}, $Rio$: \textit{Rio de Janeiro, Brazil}, $Lis$: \textit{Lisbon, Portugal}, $Por$: \textit{Porto, Portugal})}
\label{T1}
\end{table}

\section{System Architecture and Implementation \label{sec2}}
\subsection{Architecture \label{architecture}}
\begin{figure}[!t]
\centering
\includegraphics[scale=0.27]{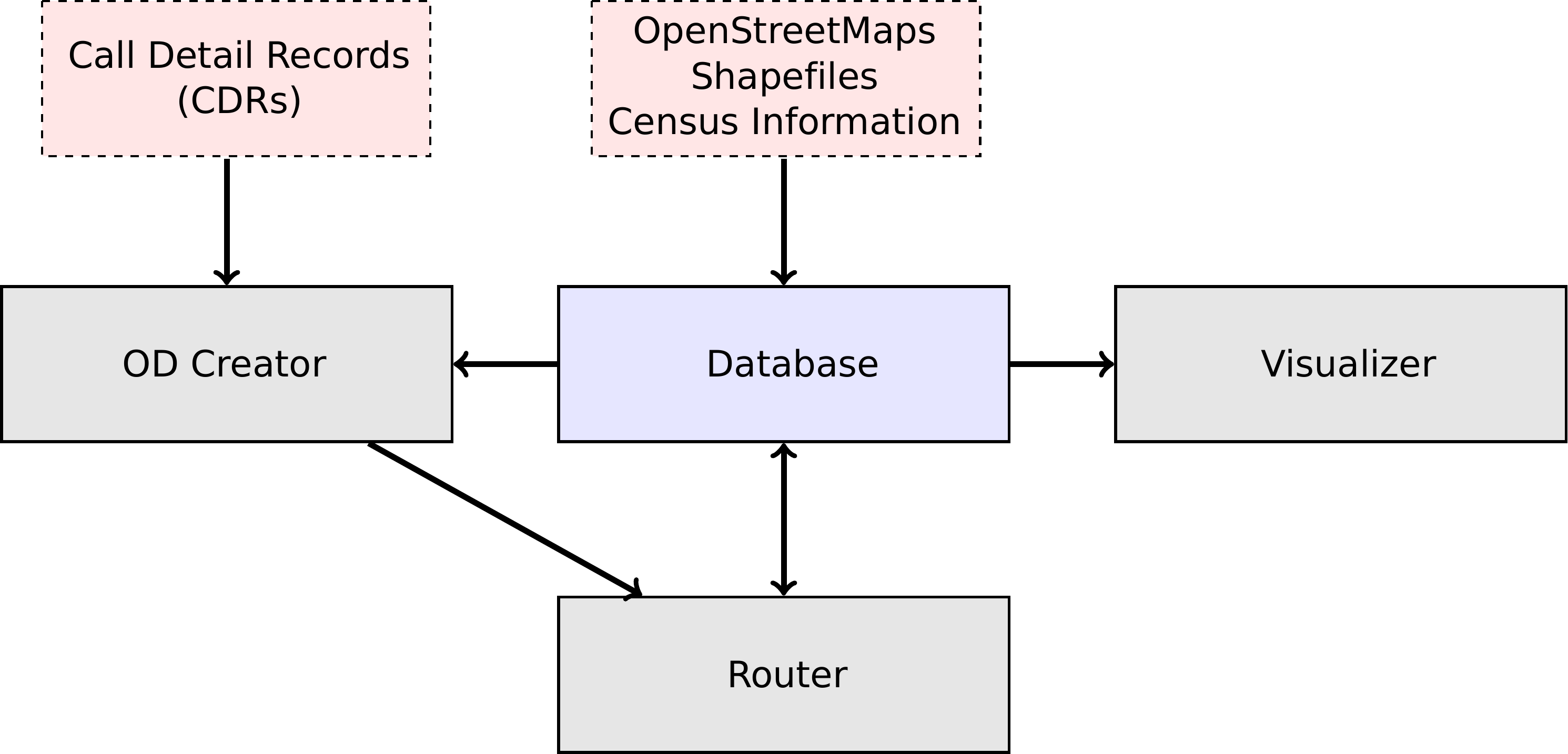}
\caption{A flowchart of the system architecture.}
\label{sysarch}
\end{figure}

Integration of the data sources described above is the critical component of this project. Our system must be flexible enough to handle different regions of the globe which may have different data availability and quality, but also efficient enough to analyze massive amounts of data in a reasonable amount of time. We would also like our system to be modular, so that components can be updated easily as new technologies become available or alternative algorithms need to be implemented. For example, GPUs are increasingly being utilized in data analyses in massively parallel environments that  significantly reduce the time needed to estimate location transition probabilities. Routing algorithms, however, do not realize these gains as will be explained later. We would therefore like our system to work with any combination of transition estimates and routing systems.  One final objective in building this modular system was to make results easily accessible to users, requiring that our system be easy to integrate with online platforms for visualization and access.  To satisfy these constraints, we propose the system architecture depicted in Figure~\ref{sysarch}.

A relational database is used to store road network and census information for every city in a standard format.  This database is accessed by tools that estimate the number of flows between origins and destinations (ODs) and algorithms that assign traffic using various route choice algorithms.  From these route assignments, road usage metrics are computed and the networks are updated back to the database.  An API is then created to power an interactive visualization platform.

\vspace{1cm}

\subsection{Creating and storing road networks \label{roadnets}}
We use a Postgres relational database to store census data and road networks, as both our routing algorithms and visualization platform depend on this information. The open source spatial extension PostGIS is used to store geographic attributes of road network and census. The database not only supports most standard geographic data formats, but also imposes schema standards through the relational model that make downstream components more efficient and interoperable. As long as a user provides a road network with the required attributes, whether it be from a local municipality or scrapped from another source, we can be sure that the router has all the information necessary. Additionally, Postgres and PostGIS connections are also supported by most GIS platforms and can connect to geoservers and output data in formats suitable for interactive visualization.

While the platform supports road networks supplied by local municipalities in the form of shapefiles, we have implemented a parser to construct routable road networks from OpenStreetMap (OSM) data due to its global availability.  While many geographic features are available within OSM data, we focus our attention on {\it node} and {\it way} elements relevant to transportation networks. Nodes represent points in space that can refer to anything from a shop to a road intersection, while ways contain a list of references to nodes that are chained together to form a line. In our context, relevant ways are those used by cars.  Ways and nodes may also contain a number of tags to denote attributes such as ``number of lanes" or ``speed limit".  Many roads, however, do not include the whole set of attributes necessary for accurate routing. For example, city roads often lack speed limit information required to estimate the time cost, which in turn is used to find shortest paths based on total travel time. To infer this missing data, our system supports the creation of user-defined mappings between highway types and road properties.  Ways tagged as ``motorways" are generally major highways and have a speed-limit of 55 mph in the Boston area.  They tend to have 3 lanes in each direction.  ``Residential" roads, on the other hand, have a speed-limit of 25mph and 1 lane in each direction.  Each road segment is also given a capacity based on formulas suggested by the US Federal Highway Administration.  Using these mappings, we parse the OSM xml data to create a routable, directed road graph with all properties required to estimate realistic costs driving down any given road.

We implement two additional cleaning steps to improve efficiency. The first filters out irrelevant residential roads. These small local roads are filtered from our network, as they are not central to the congestion problem, yet tend to increase computation time significantly. Finally, in OSM data, a node object can refer to many things, for example an actual intersection or simply a vertex on a curve used to draw a turn.  The latter case results in a network node with only one incoming and one outgoing edge (assuming U-turns are not allowed). These nodes are superficial and increase network size and routing algorithm run times needlessly.  We collapse networks by removing these nodes from the network and only connecting true intersections, keeping the geographic coordinates of the nodes so that link costs still do reflect actual geographic length of roads rather than straight line distances. The parsed and cleaned edges are then loaded into the Postgres database, preserving attributes and geometry.

\subsection{OD Generation \label{odgeneration}}

\begin{figure*}
\centering
\includegraphics[scale=0.35]{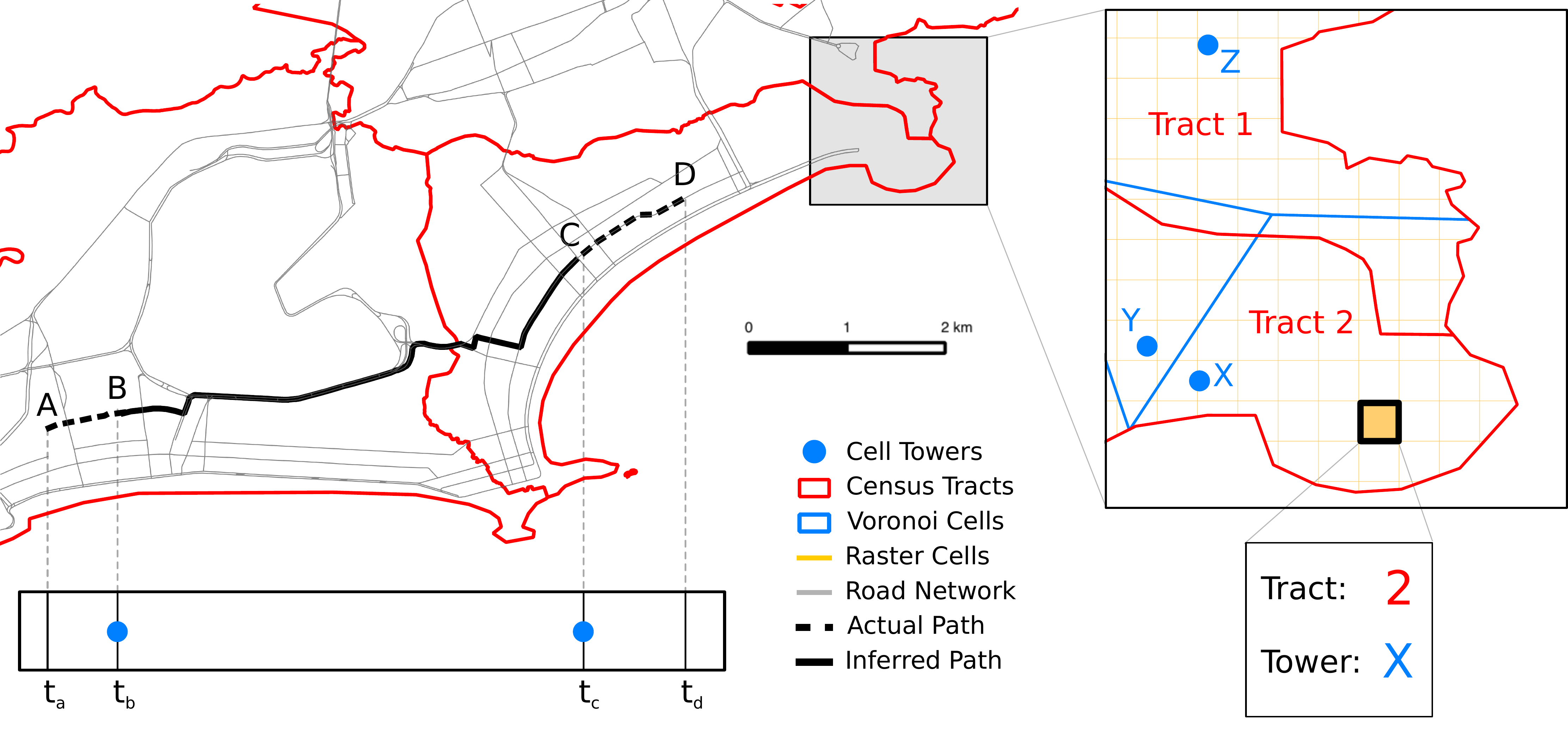}
\caption{A schematic of the OD creation through the extraction of trips, and the rasterization process.}
\label{od_fig}
\end{figure*}

\begin{algorithm}[!t ]
\caption{Estimating OD Matrices from CDRs}
\begin{algorithmic}
 \STATE $N_o^{users}\leftarrow 0$ for each location $o$
 \STATE $OD_{o,d}\leftarrow 0$ for all location pairs $o,d$

 \FORALL{ users $u$ | $n_{min} < u.numCalls < n_{max}$}
 	\STATE $u.calls \leftarrow $ vector of calls of $u$, sorted by time
 	\STATE $u.home \leftarrow$ most visited location betw. 9pm and 7am
    \STATE $N_{u.home}^{users} \leftarrow N_{u.home}^{users} + 1$

 	\STATE $o \leftarrow inPolygon(u.calls[0].location)$$^{*}$
 	\STATE $t_{o} \leftarrow u.calls[0].time$
    \FOR{$i=1$ to $i=u.calls.length()$}
        \STATE $d \leftarrow inPolygon(u.calls[i].location)$
    	\STATE $t_{d} \leftarrow u.calls[i].time$
    	\IF{$t_{o}-t_{d}<t_{diff}$ \AND $t_{start} < t_{o},t_{d} < t_{end}$ }
        	\STATE $OD_{o,d} \leftarrow OD_{o,d} + 1$
        \ENDIF
        \STATE $o \leftarrow d$; $t_{o} \leftarrow t_{d}$
    \ENDFOR
 \ENDFOR
 \\\hrulefill
 \STATE $totalTrips \leftarrow \sum_{o,d}OD_{o,d}$
 \STATE $VUR_o,POP_o$ loaded from database
 \STATE $W \leftarrow \sum_oPOP_o\cdot 4\cdot 1/24$, total hourly trips
 \FOR{ $o,d$ in $OD$}
 	\STATE $OD_{o,d} \leftarrow W\cdot VUR_o \cdot\frac{POP_o}{N_{o}^{users}}\cdot    \frac{OD_{o,d}}{totalTrips} $
 \ENDFOR
 \\\hrulefill
 \STATE $^{*}$ inPolygon(b) returns the census tract from which the call was made.
 
\end{algorithmic}
\label{od_algorithm}

\end{algorithm}

Our goal in the data mining process is to produce estimates for the number of individuals traveling by car between any origin-destination (OD) location pair in the city. Traditional methods of creating OD pairs generally consist of two steps: trip generation and trip distribution. These steps can be summarized as follows. The number of trips originating in a given place may be estimated based on a mix of amenities, businesses, or residents there or by activity based models, which use travel diaries and surveys to estimate the number of trips required to satisfy an individuals needs. After trips are generated at origins, they are assigned to destinations, often using physically inspired gravity or radiation models \cite{ortuzar1994modelling, simini2012universal}. While considerable work has been done to make these estimations accurate using complex methods of calibration, new data collected from mobile devices and applications provide alternative means to count OD flows in a more direct fashion. 

However, with new data sources come new challenges. Arguably the most crucial challenge lies in the fact that trips can only be observed when a user interacts with his or her phone. This results in non-uniform sampling frequencies, an issue further amplified by the penetration rates of mobile phone usage, market shares of the carriers involved, and usage patterns, among many other factors.

To convert raw call detail records into reliable OD flows, we implement the procedure elaborated in Algorithm~\ref{od_algorithm}. First, we count raw trips observed from mobile phone users in order to estimate transition probabilities between locations.  This involves measuring the number of calls made by each user, $u.numCalls$, and inferring their home location $u.home$ from the place visited the most at night. By counting the number of users in our dataset that we deem to live in a certain area $o$, $N^{users}_{o}$, we can infer the population of phone subscribers at each place. Because CDR data typically comes in one of two formats, with locations denoted by tower IDs or triangulated latitude and longitudes, we have two different versions of this procedure which are structurally identical. In both cases, a trip is marked by two consecutive calls $t_{o}$ and $t_{d}$, occurring at two different locations within a specified time window, $t_{diff}$. For the purposes of this study, we are interested in morning commute periods and set $t_{start} = 6am$ and $t_{end} = 9am$ and $t_{diff} = 1 hr$. In the case of tower based CDRs, trips between two towers are assigned to census tracts covered by the Voronoi cell served by that tower, in proportion to the area of intersection of the Voronoi and the census tract polygons. Coordinate based CDRs are mapped directly into the census tract containing the calls.

With billions of phone calls to consider, the spatial join procedure has a large impact on computation speed.  Vectorized versions of an algorithm to determine if a point is contained within a polygon are algorithmically simple, but computationally slow as a naive implementation checks if the point belongs in each possible census tract.  Instead, we utilize a far faster approach.  We \textit{rasterize} the census polygons by overlaying a uniform grid and labeling each cell with the ID of the census tract that covers most of its area. This creates some error for cells that lie on the boundary of tracts, but this can be minimized by choosing a small enough raster and is often far smaller than the error in localization of phones.  In essence, the raster creates a convenient lookup table that can be stored in memory so that each coordinate pair can be quickly placed within a polygon. Figure~\ref{od_fig} depicts the rasterization process in the case of tower based CDR data.

Figure~\ref{od_fig} also exhibits a common scenario in Rio, where the CDR data is in tower resolution.  A user leaves a point A at time $t_A$ $(A, t_A)$ and initiates an event on a mobile device at point B at time $t_B$ $(B,t_B)$.  Later, at $(C, t_C)$, the user initiates another mobile event and eventually finishes his trip at $(D,t_D)$.  We are unable to observe $(A,t_A)$ or $(D,t_D)$, but count the trip between $B$ and $C$ so long as $t_B - t_C < t_{diff}$ and $t_{start} < t_B,t_C < t_{end}$. This scenario captures the transient nature of the OD information that can be obtained through CDR data. Although this issue appears inherent considering the frequency of phone calls, it will become less disruptive as both the spatial and the temporal resolution of CDRs improve in parallel with telecommunications technology.

The result of this procedure is a list of pairs of OD census tracts and the counted trips between them. However, due to differences in market share, mobile phone penetration, and mode choice by users, these counts do not reflect the actual number of vehicle trips taken.  To correct for these biases, we scale by two quantities. First, we multiply the number of trips between each pair by the ratio of census tract population to mobile phone users identified to be living in the tract of trip origin, $\sfrac{POP_o}{N_{o}^{users}}$. This provides an upscaled estimate for the total number of trips observed between each census tract pair. 

The second scaling is required by the fact that some of these trips may not be completed via a car as is often the case in downtown regions with good public transit or close proximity for walking. This branching of trips into various modes of travel is referred to as the mode choice step in transportation planning. In order to account for this, we multiply the upscaled total trip count by the vehicle usage rate of individuals living in the origin tract, $VUR_{o}$. We are left with an estimate for the number of vehicle trips observed by mobile devices between each pair of census tracts in the city.  We then convert these counts to transition probabilities by dividing the flow between each pair by the total number of upscaled trips observed $\sfrac{OD_{o,d}}{totalTrips}$.

While CDR data is well suited for inferring transition probabilities between locations, there is a need scale these probabilities to obtain total travel demand across the area. Therefore the next step of OD creation is to convert transition probabilities to absolute numbers of trips which allows use to make better use of our data by avoiding agent or gravity-based models in our aim to obtain realistic trip distributions. In doing this, we multiply the transition probability matrix by $W$, equal to the total number of trips taken across the entire city in a typical hour during the morning commute period, computed by the total population of the region times the average number of trips taken per person on a given day. Using constants supplied in \cite{wang2012understanding}, we assume 4 trips per person on average based on estimates used in the US cities. Then we account for the fraction of these trips taken during a typical morning rush hour by scaling the total number by $\sfrac{1.5}{24}$. This finally gives us a rough estimate of the total number of trips taken between any two census tracts in the city during a typical hour on a weekday morning. In the final step of the algorithm, we retrieve the location of all intersections from the database, join these intersection to census tracts, and assign them trips uniformly at random until all tract-to-tract trips have been assigned to an intersection-to-intersection pair.

\subsection{Trip Assignment \label{assignment}}
Having estimated intersection to intersection OD pairs, our next task is to efficiently route hundreds of thousands of trips through the road network \cite{bast2007fast}.  We develop a set of tools built in C++ to perform large scale routing and traffic assignment using multithreading for parallelization speedups.  First, the user specifies a road network table from the database which is downloaded and used to create a directed graph object using the Boost Graph Library.  We can then compute shortest paths based on a user defined cost (in this case travel time on road segments). We choose the A* algorithm among the wide range of shortest path algorithms, as it's widely used in routing on geographic networks for its flexibility and efficiency. The A* algorithm implements a \textit{best-first-search} using a specified heuristic function to explore more promising paths first.  The euclidian distance between nodes provides an intuitive heuristic that ensures optimal solutions are found. Because our aim is to estimate road usage across the entire network, it may seem more efficient to use Dijkstra's algorithm to pre-compute all shortest paths in the network as opposed to calculating single paths independently. However, this approach becomes sub-optimal when an incremental traffic assignment is to be implemented, as congestion costs will be iteratively updated to properly model the traffic flow-travel time relationships.  Moreover, the A* algorithm may be a better proxy for human decision making, as it uses a heuristic function to make decisions with only local information and does not require global knowledge of the network.

\begin{figure}
\centering
\includegraphics[scale=0.5, trim=0mm 5mm 0mm 0mm]{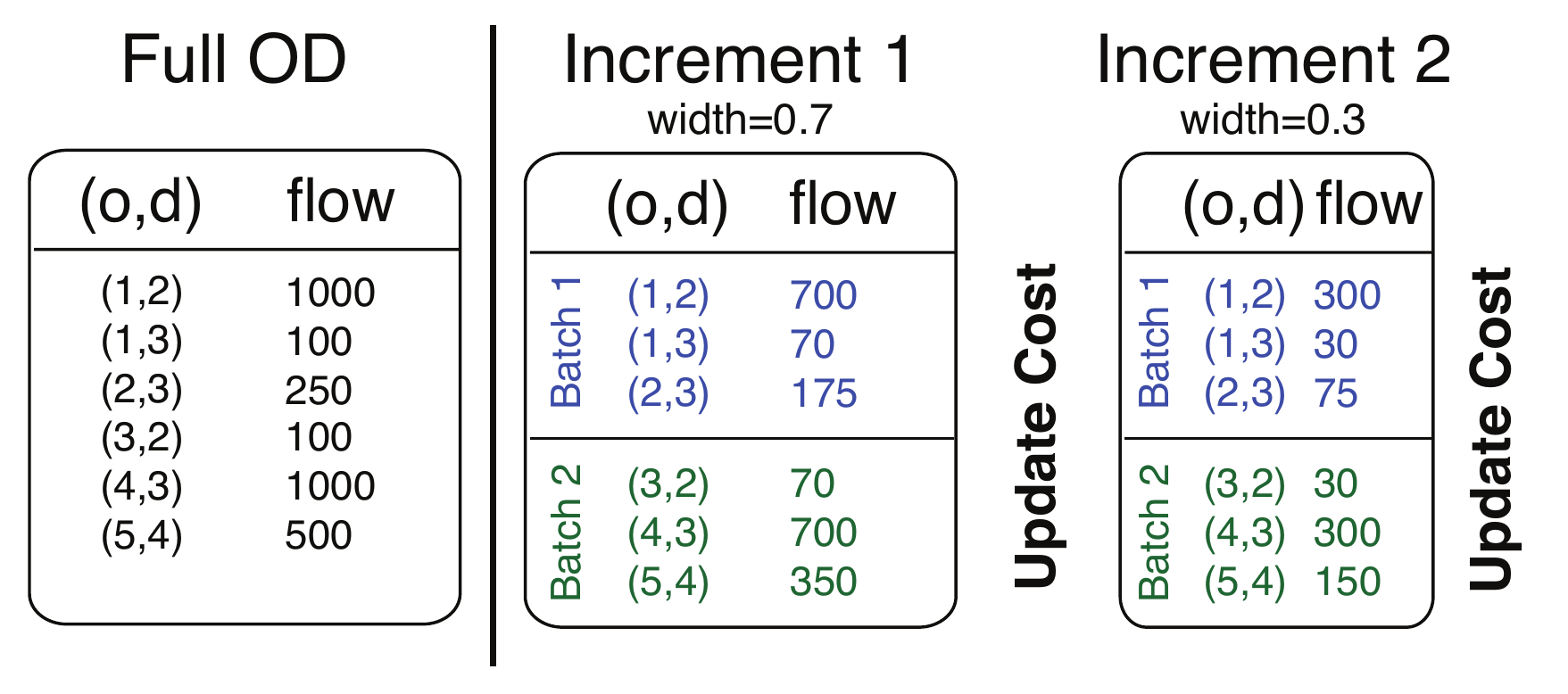}
\caption{Our efficient implementation of the incremental traffic assignment (ITA) model. A sample OD matrix is divided into two increments and then parallelized into two batches each. \label{ita_fig}}
\end{figure}

On most city roads, free-flow speeds are rarely achieved due to congestion. As a result, traffic patterns may significantly change the time costs associated with using a particular route. To ensure that these feedback mechanisms are accounted for in road usage modeling, Incremental Traffic Assignment algorithms have been developed \cite{vaze2009calibration, daganzo1977stochastic, ortuzar1994modelling, patriksson1994traffic}. A simplified schematic explaining the procedure can be seen in Figure \ref{ita_fig}. These algorithms assign trips in a series of increments and update the costs of edges in the network based on the number of vehicles that were previously assigned to that road.  For example, the first increment assigns 40\% of trips for each pair assuming each driver experiences free-flow speeds. The travel time cost associated with every road segment is then adjusted based on how many drivers were assigned to that road and the total number of cars a road can accommodate in unit time.  The next 30\% of drivers are then routed in the updated conditions.  This process is repeated until all users have been assigned a route.  

A common metric used in determining the travel time associated with a specific flow level is the ratio between the number of cars actually using a road (volume) and it's maximum flow capacity (volume-over-capacity or VOC).  At low VOCs, drivers enjoy large spaces between cars and can safely travel at free-flow speeds.  As roads become congested and VOC increases, drivers are forced to slow down to insure they have adequate time to react. Based on the volume-over-capacity (VOC) for each road, costs are updated according to Eq. \ref{E_bpr}, where $\alpha=0.15, \ \beta=4$ are used per guidelines set by the Bureau of Public Roads\footnote{Travel Demand Modeling with TransCAD 5.0, User's Guide (Caliper., 2008).}. 
\begin{equation}
\label{E_bpr}
t_{current} = t_{free flow}\cdot (1 + \alpha (VOC)^\beta)
\end{equation}

Though increments must be routed in serial, all routes discovered within an increment are independent.  To speed up the routing process, we divide all trips in an increment into batches and send these batches to different threads for parallel computation.  Because the road network remains fixed in each increment, we only need to store a single graph object shared by all threads.  When a shortest path is found, we walk that path and increment counts of the number of vehicles that were assigned to each road and sum the counts from all batches after the increment has finished.  We also keep track of the origin and destination census tracts of the assigned vehicles in a bipartite graph for later analysis. After all trips have been routed, we compute final VOC ratios and other metrics of each segment and update these values in the database so they can be used for other applications or visualization. The complete routing procedure is summed up in Algorithm~\ref{routing_algorithm}.

\begin{algorithm}
\caption{Trip Assignment}
\begin{algorithmic}
 \STATE $G\leftarrow$ road network loaded from database
 \STATE $B \leftarrow$ a bipartite network containing roads and census tracts
 \STATE $OD\leftarrow$ loaded from file
 \STATE $incrSize \leftarrow$ vector of supplied by user at run time
 \STATE $nBatches \leftarrow$ number of threads to use
 
 \FOR{ $i=0$ \TO $i<incrSize.length$}
 	\FOR {$b=0$ \TO $b<nBatches$}
    	\STATE start new thread
        \STATE $batch \leftarrow OD.getBatch(b) $ $^{*}$
        \FORALL {$o,d$ pairs in $batch$}
        	\STATE $flow \leftarrow OD_{o,d}\cdot incrSize[i]$
        	\STATE $route \leftarrow A^*(o,d,G)$ $^{+}$
            \FORALL {segments $s$ in $route$ }
            	\STATE $s.volume \leftarrow s.volume+flow$
                \STATE $B_{s\rightarrow o} \leftarrow B_{s\rightarrow o} + flow$
            \ENDFOR
        \ENDFOR
	\ENDFOR
    \STATE wait for all threads to finish
    \FOR {segment $s$ in $G$}
    	\STATE $s.cost \leftarrow s.freeFlowTime\cdot(1+\alpha(\frac{s.volume}{s.capacity})^\beta)$
        
    \ENDFOR
 \ENDFOR
 \\\hrulefill
 \STATE $^{*}$ getBatch(b) returns only the subset of OD pairs pertaining to a batch
 \STATE $^{+}$ A*(o,d,G) returns the shortest path between $o$ and $d$ if a path exists 
\end{algorithmic}
\label{routing_algorithm}
\end{algorithm}

\section{Results}

\subsection{Road Network Performance\label{vvoc}}

\begin{figure}
\centering
\includegraphics[scale=0.45, trim=7mm 10mm 0mm 10mm]{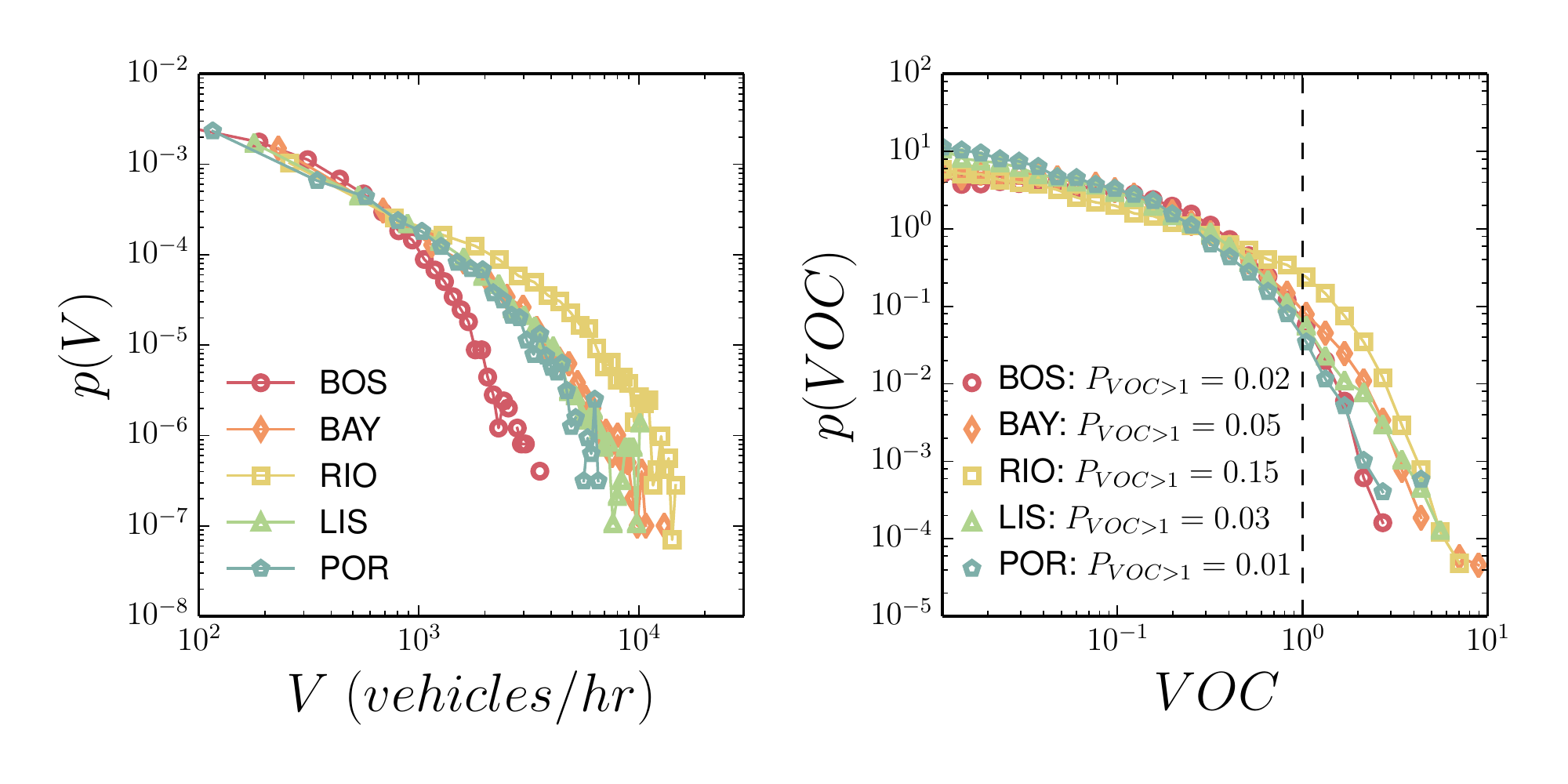}
\caption{Distributions of travel volume assigned to a road and the volume-over-capacity (VOC) ratio for the five cities. The values presented in the legend refers to the fraction of road segments with $VOC>1$. \label{voc_fig}}
\end{figure}

One output of our procedure is estimated travel volume on every road segment. We produced the distributions of volumes on roads, along with $VOC$s in Figure \ref{voc_fig}. Our results suggest qualitatively similarly distributed volumes and VOCs for our five subject cities. Moreover, our findings are consistent with general congestion studies that identify Rio de Janeiro as one of the most congested cities in the world and the San Francisco Bay Area not far behind. The legend demonstrates the ratio of road segments that have $VOC>1$, or in other words roads that are suffering from delays. This further strengthens Rio's position among the analyzed cities as the most congested, as $15\%$ of road segments suffer from congestion in the morning rush, as opposed to other cities where this number does not exceed $5\%$. It can be observed that smaller municipalities such as Boston and Porto have fewer problems with congestion.

\subsection{Bipartite Road Usage Graph \label{bipartite}}
\begin{figure}
\centering
\includegraphics[scale=0.3]{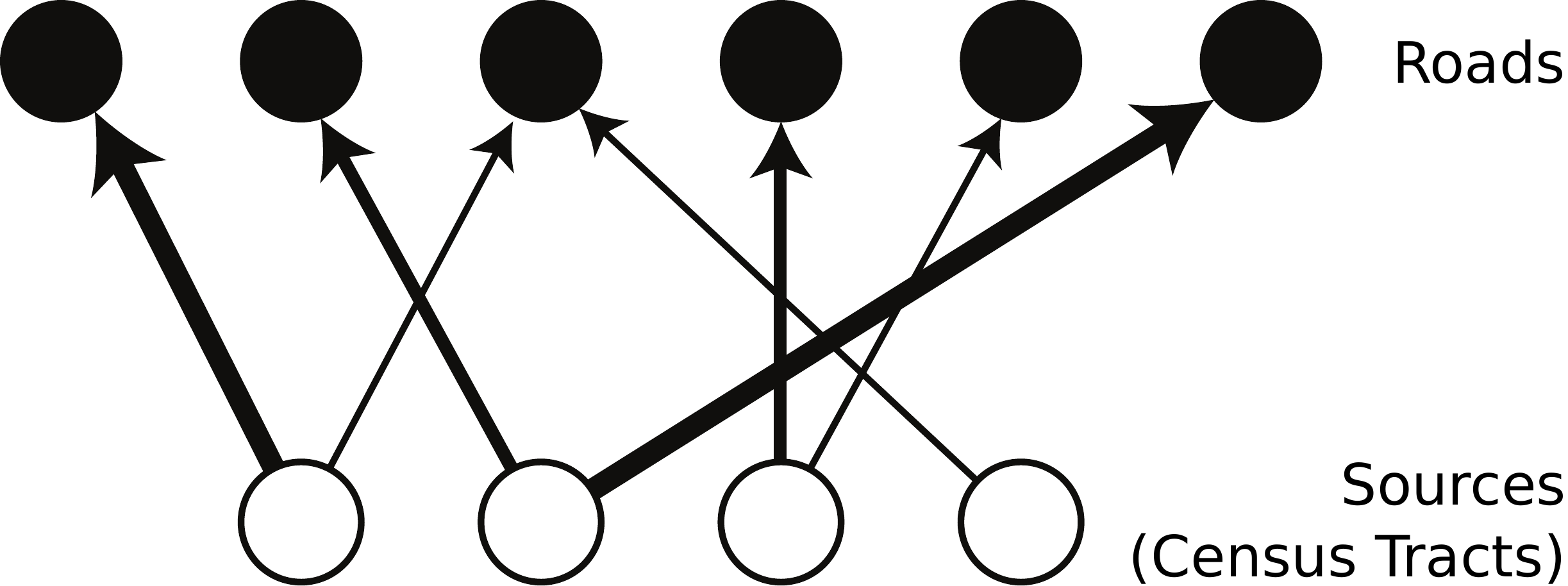}
\caption{A graphical representation of the bipartite network of roads and sources (census tracts), with edge sizes mapping the number of users using the connected road in their individual routes. \label{bipartite_fig}}
\end{figure}

In addition to measuring physical network properties of roads, our system also enables detailed analysis of individual road segments and neighborhoods within a city.  Every time a route between two location is assigned, we traverse the path and keep a record of how many trips from each driver source (census tract) used each road \cite{wang2012understanding}.  This record is then used to construct a bipartite graph containing two types of nodes: road segments and driver sources, as shown in Figure \ref{bipartite_fig}.  Roads are connected to driver sources that contribute traffic to that segment and census tracts are connected to roads that are used by people who live here.  

\begin{equation}
\label{E_bipartite}
k_s^{road} = \sum_o A_{o\rightarrow s}, \ k_o^{source}=\sum_s A_{o\rightarrow s}
\end{equation}

\[
 A_{o \rightarrow s} =
  \begin{cases}
   1, & \text{if vehicles from tract $o$ use road $s$} \\
   0, & \text{otherwise.}
  \end{cases}
\]

\begin{figure}[!t]
\centering
\includegraphics[scale=0.7]{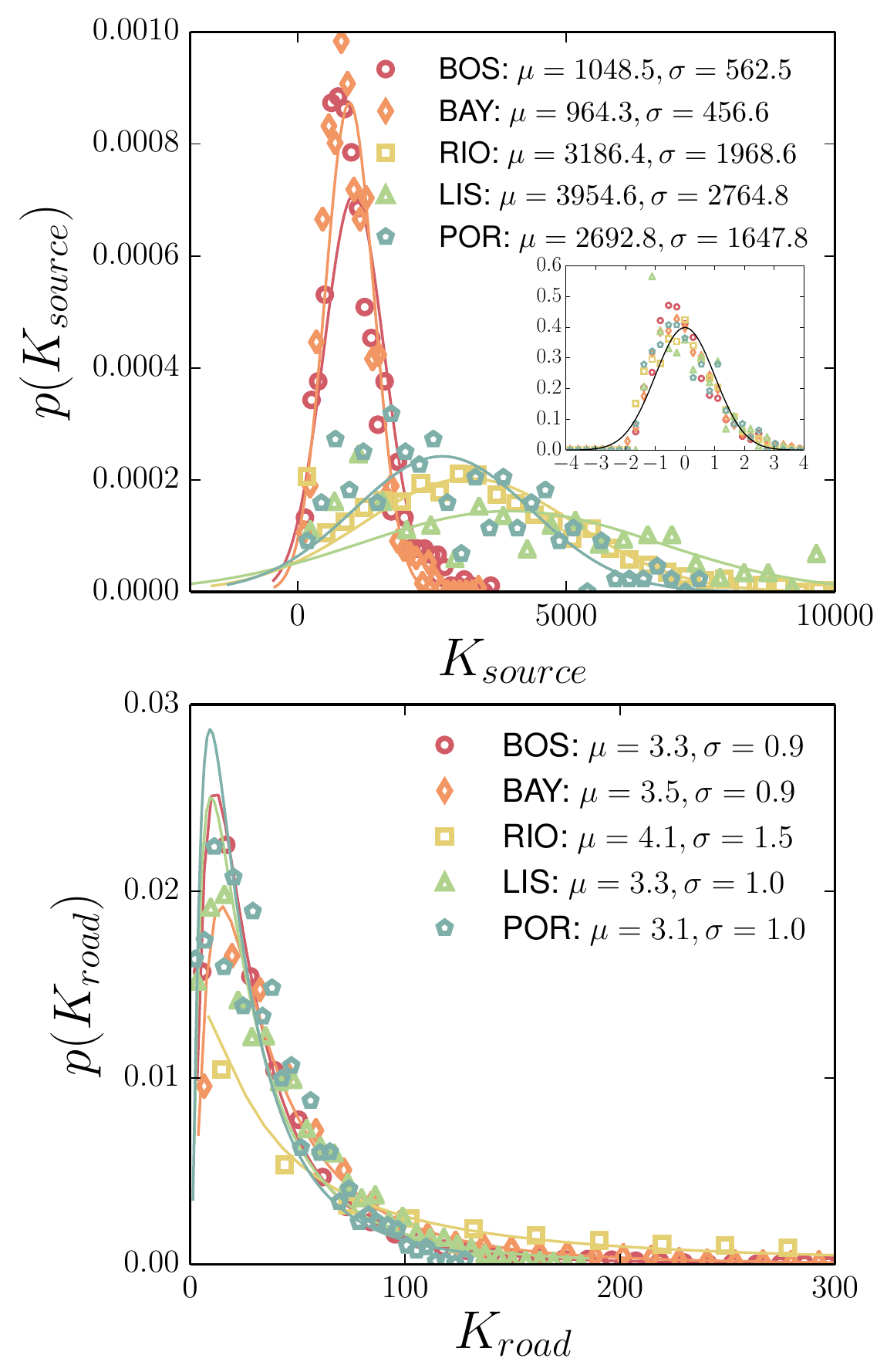}
\caption{Distributions of $K_{road}$ and $K_{source}$ for the five cities. Inset: The unitized collapsed normal distribution for $K_{source}$. The associated exponents are expressed in the legend.  \label{kroad_fig}}
\end{figure}

We then examine the degree distributions of roads and census tracts (\ref{E_bipartite}) in this bipartite graph to reveal patterns of road usage in Figure \ref{kroad_fig}.  The number of roads used by residents of a given location in a city is normally distributed.  The mean and standard deviation differs depending on the spatial resolution of the driver source and detail of the road network, which is the case for Rio de Janeiro and its significantly larger road network, but all cities collapse onto a standard normal.  On the other hand, the number of driver sources contributing traffic to a given road segment is well fit by a lognormal distribution for each city, suggesting that most roads are \textit{local} in that they serve only a few locations, while a few roads in the tail of the distribution serve as connectors for the whole city.

Bipartite graphs allows us to augment visualizations of congestion maps in two ways.  The first focuses on a single road segment.  For example, when we identify a segment of a highway that becomes highly congested with traffic jams each day, we can easily query the bipartite graph to obtain a list of census tracts where drivers sitting in that traffic jam are coming from and where they are going to.  The census tract nodes can also be given attributes from containing any demographic data a user wishes.  With this information, it is possible to identify leverage points where policy makers can offer alternatives to these individuals or even power applications such as car sharing, by notifying drivers that others sharing the same road may be going to and from the same places.  Moreover, businesses considering products or services based who may be driving by or near different locations may find value in these detailed breakdowns.

Rather than selecting a road segment node, we may also select a single census tract, and check it's neighbors to construct a list of all roads used by individuals moving to or from that location.  For example, for a given neighborhood in a city we can identify all major arteries that serve that local population.  This information provides a detailed look at a central location based on how much road usage it induces.  Moreover, geographic accessibility, critical to many socio-economic outcomes, can now be measured in locations that were previously understudied.

\subsection{Road Classification\label{roadclass}}
Comparing the topological properties of roads in the physical network to the bipartite usage graph provides insights into their role in the transportation system.  Edge betweenness centrality \cite{newman2005measure} captures the importance of a road by counting how many shortest paths between any two locations $\sigma_{od}$ must pass through that edge $\sigma_{od}(e)$ (Eq. \ref{E_betweenness}).  While this measure captures some aspects of importance, it treats all potential paths as equally likely and tends to be biased towards geographically central links.  The degree of a road in the bipartite usage graph reflects the number of locations in the city that actually rely on that road because trips were assigned there from actual travel demand.  With these two metrics, betweenness centrality and a roads degree in the usage network, we can classify the role of a road in the cities transportation network.

\begin{figure}[!t]
\centering
\includegraphics[scale=0.4,trim=0mm 0mm 0mm 0mm]{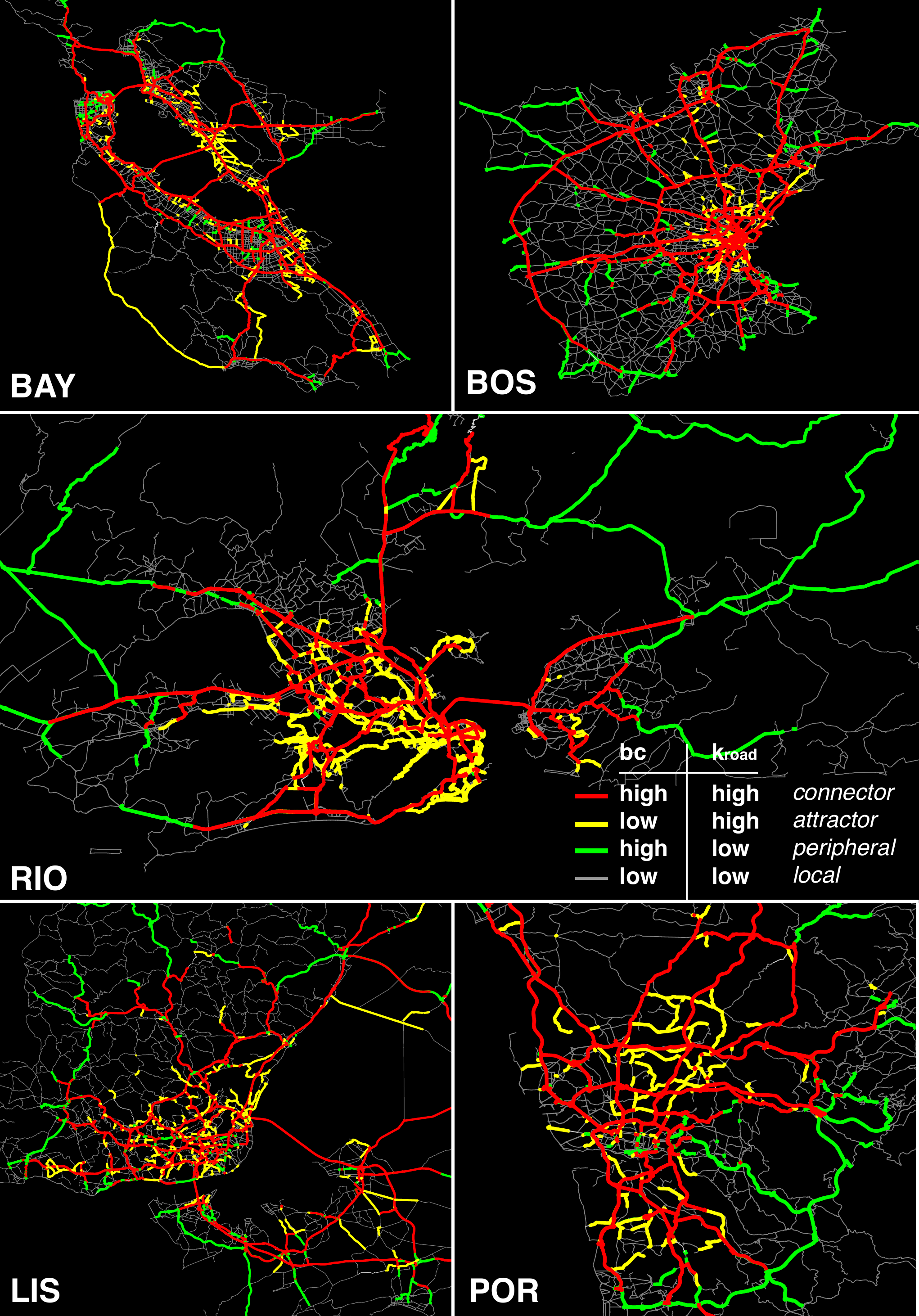}
\caption{Maps depicting the proposed road classification, summarized in the legend, for the five subject cities.\label{roadclass_fig}}
\end{figure}

\begin{equation}
\label{E_betweenness}
bc_s = \sum_{o,d}\frac{\sigma_{od}(s)}{\sigma_{od}} 
\end{equation}

A simple classifier divides the betweenness usage degree space into four quadrants surrounding the point representing the 75th percentile for betweenness centrality and usage degree.  Roads with betweenness and usage degree above the 75th percentile are both physical connectors and are used by large portions of the region.  These roads tend to be bridges or urban rings.  Roads with low betweenness, but high usage degree are attractors, receiving a higher proportion of trips than would be expected assuming uniform demand.  Roads with high betweenness and low usage are physical connectors and serve an important purpose geographically, but may not be utilized by actual demand.  Other roads, with low betweenness and low usage are local roads and primarily serve populations living and working nearby.  Figure \ref{roadclass_fig} shows each road according to this classification.

\begin{figure}[!h]
\centering
\includegraphics[scale=0.16, trim=0mm 0mm 0mm 0mm]{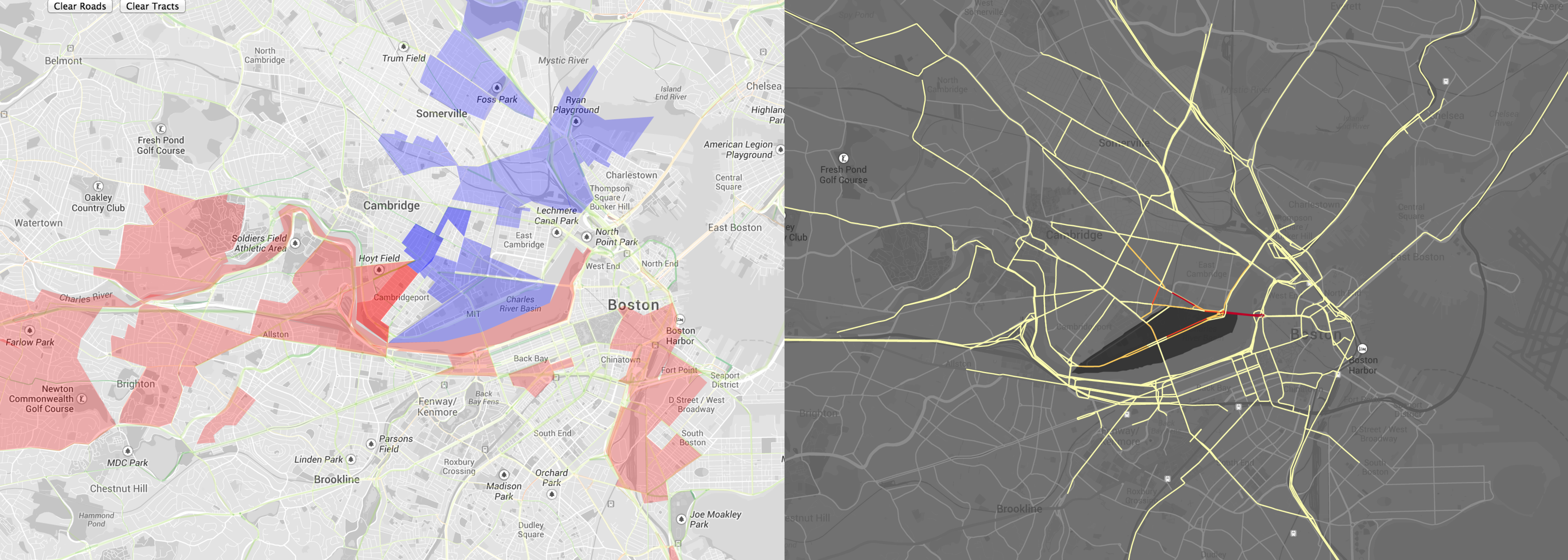}
\caption{Two screen images from the visualization platform. (a) The trip producing (red) and trip attracting (blue) census tracts using Cambridge St., crossing the Charles River in Boston. (b) Roads used by trips generated at the census tract including MIT. \label{vis_fig}}
\end{figure}

\vspace{1cm}

\subsection{Benchmarking \label{benchmark}}

\begin{table}
\centering
\begin{tabular}{lccccc}
 & \multicolumn{5}{c}{City}  \\
\cline{2-6}
 & Bos & Bay & Rio & Lis & Porto \\
\hline
OD pairs (1000s) & 305 & 398 & 850 & 340 & 154\\
Time / incr. (s) & 5.6	& 10.3 & 101.3 & 23.4 & 7.9\\
Time / route (ms) & 0.018	& 0.026 & 0.12 & 0.067 & .051 \\
\hline
\end{tabular}
\caption{Benchmarking routing performance for each city.  All tests use 10 threads. \label{bm_table}}
\end{table}

%\begin{figure}
%\centering
%\includegraphics[scale=0.5]{benchmarking.pdf}
%\caption{Average time spent to route a single origin-destination pair with the number of threads involved. \label{bm_fig}}
%\end{figure}

We perform these analyses on a server containing 48GB of ram and 12 cores between 2 Intel Xenon processors. Table \ref{bm_table} shows routing performance for different cities. We test for performance using different number of threads and a 600\% speedup over serial execution was observed using 10 threads. 

\subsection{Visualization \label{visualization}}
To help make these results accessible to consumers and policymakers, we build an interactive web visualization to explore road usage patterns in each city.  Most GIS platforms can connect directly PostGIS databases to visualize and analyze road networks with our estimated usage characteristics.  While these platforms are preferred by advanced users familiar with GIS data, they are opaque to many consumers who may benefit from more detailed information on road usage. A simple API is implemented to query the database and generate standard GeoJSON objects containing geographic information on roads as well as computed metrics such as level of service.  We also implement queries to answer questions such as ``What are all the census tracts used by drivers on a particular road?" or "What are all roads used by a given location in the city?".  These data are then parsed and displayed on interactive maps using any of the available online mapping APIs and D3js allowing users, with functionality that enables one to select individual roads and areas. Two screen images of this system is shown in Figure \ref{vis_fig}.

\vspace{1cm}
\section{Conclusions \label{conclusion}}

We have demonstrated a system to analyze massive spatiotemporal datasets to generate new insights into the interaction between a city's inhabitants and its transportation infrastructure. While we hope future work continues to refine and calibrate these inference algorithms, we have shown their promise by estimating travel demand, mapping  demand onto roads, and exploring the resulting patterns of level of service and usage in five cities. We devised metrics by which one can assess the role of road segments or census tracts through the bipartite network of connections between the two and classified roads based on the topology this usage and it's physical topology. Finally, we presented a visualization platform allowing users to explore these traffic patterns from the web. We believe our framework not only lays a foundation on which numerous research questions concerning human mobility, cities, traffic and infrastructure networks can be better answered, but also is suitable for use by decision makers, including departments of transportation, municipalities and the units government involved. %There's also further premise in the use of similar systems as they can generate information of value through developing a better understanding of regions, people and systems, or simply the complex system that is the city. 

\section{Acknowledgments}
This work was partially funded by the BMW-MIT collaboration under the supervision of PI Mark Leach\footnote{mark.leach@bmw.de}, the World Bank-HuMNet collaboration agreement under the supervision of PI Shomik Mehndiratta\footnote{smehndiratta@worldbank.org} and the Center for Complex Engineering Systems (CCES) at KACST under the co-direction of Anas Alfaris\footnote{anas@mit.edu}. We thank Lauren Alexander and Pu Wang for technical support, Nelson F. F. Ebecken for support with data, the Rio de Janeiro State Agency (FAPERJ) for the grant on this project and the Rio City Hall for the support and the data they have provided. Our work was also supported, in part, by the UPS Center for Transportation and Logistics Graduate Research Fellowship awarded to Serdar \c{C}olak and by the National Science Foundation Graduate Research Fellowship awarded to Jameson L. Toole.

\bibliographystyle{abbrv}
%\bibliographystyle{acm}
%\bibliographystyle{unsrt}
%\bibliography{sigproc}  % sigproc.bib is the name of the Bibliography in this case

\end{document}